\documentclass[pra,aps,amsmath,amssymb,showkeys,showpacs]{revtex4}
\newcommand{\hh}{\mathcal{H}}
\newcommand{\lnp}{\mathcal{L}}

\newcommand{\lsp}{\mathcal{L}_{+}}
\newcommand{\bro}{\boldsymbol{\rho}}
\newcommand{\wbro}{\widetilde{\boldsymbol{\rho}}}

\newcommand{\pen}{\openone}

\newcommand{\ax}{\mathsf{X}}
\newcommand{\ay}{\mathsf{Y}}
\newcommand{\fm}{\mathsf{F}}

\newcommand{\ppm}{\mathsf{P}}
\newcommand{\qpm}{\mathsf{Q}}

\newcommand{\Tr}{\mathrm{Tr}}
\newcommand{\cla}{\mathcal{A}}
\newcommand{\clb}{\mathcal{B}}

\newcommand{\ppc}{\mathcal{P}}
\newcommand{\qpc}{\mathcal{Q}}
\newcommand{\cmb}{\mathbb{B}}
\newcommand{\mpb}{\mathbb{P}}
\newcommand{\mqb}{\mathbb{Q}}

\begin{document}
\clearpage
\preprint{}

\title{On uncertainty relations and entanglement detection with mutually unbiased measurements}
\thanks{Presented at 46th Symposium on Mathematical Physics ``Information Theory \& Quantum Physics'', Toru\'{n}, Poland, June 15--17, 2014.}
\author{Alexey E. Rastegin}
\affiliation{Department of Theoretical Physics, Irkutsk State
University, Gagarin Bv. 20, Irkutsk 664003, Russia}

\begin{abstract}
We formulate some properties of a set of several mutually unbiased
measurements. These properties are used for deriving entropic
uncertainty relations. Applications of mutually unbiased
measurements in entanglement detection are also revisited. First,
we estimate from above the sum of the indices of coincidence for
several mutually unbiased measurements. Further, we derive
entropic uncertainty relations in terms of the R\'{e}nyi and
Tsallis entropies. Both the state-dependent and state-independent
formulations are obtained. Using the two sets of local mutually
unbiased measurements, a method of entanglement detection in
bipartite finite-dimensional systems may be realized. A certain
trade-off between a sensitivity of the scheme and its experimental
complexity is discussed.
\end{abstract}

\maketitle

\pagenumbering{arabic}
\setcounter{page}{1}

\section{Introduction}\label{sec1}

Complementarity and entanglement are basic concepts of quantum
theory. Heisenberg's uncertainty principle \cite{heisenberg} is
one of the most known restrictions imposed in the quantum world. A
truly non-classical character of entanglement was emphasized in
the Schr\"{o}dinger ``cat paradox'' paper \cite{cat35}. Today,
quantum properties are considered as powerful resources for
potential usage in communication and computation tasks
\cite{nielsen}. Historically, uncertainty relations were focused
on pairs of canonically conjugate variables \cite{lahti,brud11}.
Recent researches have shown that uncertainty relations give a
useful tool for studying complementarity aspects \cite{ww10}.
Quantum entanglement is used as a basic tool in quantum
parallelism, quantum cryptography, quantum dense coding, and
quantum teleportation \cite{hhhh09}.

The notion of mutually unbiased bases has much many links with
recent studies of quantum information protocols (see, e.g., the
review \cite{bz10} and references therein). Mutually unbiased
bases are also an interesting mathematical subject \cite{bz10}.
For example, the problem of maximal set of mutually unbiased bases
is still open. In general, the maximal number of MUBs in $d$
dimensions is still an open question \cite{bz10}. When $d$ is a
prime power, the answer $d+1$ is known \cite{wf89}. For other $d$,
i.e., for composite numbers, we only know that the maximal number
of MUBs does not exceed $d+1$. The author of \cite{kag14} proposed
the concept of mutually unbiased measurements. The principal
result is that a complete set of $d+1$ mutually unbiased
measurements has been built explicitly for arbitrary finite $d$
\cite{kag14}. Hence, we have come across different questions
concerning a possible usage of such measurements in quantum
information science.

The aim of the present work is to study mutually unbiased
measurements in some important respects. The paper is organized as
follows. In section \ref{sec2}, preliminary facts are reviewed. We
also prove an interesting relation between R\'{e}nyi's entropies
of three different orders. In Section \ref{sec3}, we derive an
upper bound on the sum of the indices of coincidence for a set of
mutually unbiased measurements. In Section \ref{sec4}, uncertainty
relations for an arbitrary number of such measurements are derived
in terms of R\'{e}nyi's and Tsallis' entropies. Both the
state-dependent and state-independent formulations are given.
Applications of mutually unbiased measurements in entanglement
detection are examined in Section \ref{sec5}. We will see an
evidence for trade-off between a sensitivity of the scheme and
costs for its implementation. In Section \ref{sec5}, we conclude
the paper with a summary of results.

\section{Preliminaries}\label{sec2}

In this section, we review the required material. First, some
notation for spaces and operators is introduced. We then recall
the definition of mutually unbiased measurements proposed in
\cite{kag14}. Further, we discuss the R\'{e}nyi and Tsallis
entropies which will be used as measures of uncertainties. We also
prove a relation between R\'{e}nyi's entropies of three different
orders.

Due to sensitivity of quantum states, a measurement stage is one
of central questions in quantum protocols \cite{nielsen}. Hence,
some selected types of measurements are of special interest in
quantum information processing. Mutually unbiased bases are used
in quantum state reconstruction \cite{wf89}, quantum error
correction \cite{gott96,cald97}, detection of quantum entanglement
\cite{shbah12}, and the mean king's problem \cite{vaid87,ena01}.
For arbitrary $d$, however, constructing a maximal set of mutually
unbiased bases is an open problem. One may try to fit
``unbiasedness'' with weaker conditions \cite{kag14}. In this way,
we will deal with mutually unbiased measurements.

Let $\lnp(\hh)$ be the space of linear operators on
$d$-dimensional Hilbert space $\hh$. By $\lsp(\hh)$, we denote the
set of positive semi-definite operators on $\hh$. A density
operator $\bro\in\lsp(\hh)$ is normalized by $\Tr(\bro)=1$.
For operators $\ax,\ay\in{\mathcal{L}}(\hh)$, their
Hilbert--Schmidt inner product is written as \cite{watrous1}
\begin{equation}
\langle\ax{\,},\ay\rangle_{\rm{hs}}:=\Tr(\ax^{\dagger}\ay)
\ . \label{hsdef}
\end{equation}
Let $\cla=\{|\phi_{n}\rangle\}_{n=1}^{d}$ and
$\clb=\{|\varphi_{n}\rangle\}_{n=1}^{d}$ be orthonormal bases in
$d$ dimensions. They are said to be mutually unbiased, when
\begin{equation}
\bigl|\langle\phi_{m}|\varphi_{n}\rangle\bigr|=\frac{1}{\sqrt{d}}
\ , \label{twb}
\end{equation}
for all $m,n=1,\ldots,d$. The set
$\cmb=\{\clb^{(1)},\ldots,\clb^{(M)}\}$ is a set of mutually
unbiased bases (MUBs), when each two bases from this set are
mutually unbiased. When dimensionality $d$ is not a prime power,
we do not know the maximal number of MUBs that can be constructed.
The answer is not known even for $d=6$ \cite{bz10}. In some
respects, mutually unbiased bases are connected with symmetric
informationally complete measurements, shortly SIC-POVMs
\cite{bz10}. Such measurements are also not easy to construct.

The authors of \cite{kag14} proposed a concept of mutually
unbiased measurements (MUMs). Let $\ppc=\{\ppm_{n}\}$ and
$\qpc=\{\qpm_{n}\}$ be two POVM measurements, each with $d$
elements. We assume that POVM elements satisfy
\begin{align}
& \Tr(\ppm_{n})=\Tr(\qpm_{n})=1
\ , \label{tmn1}\\
& \Tr(\ppm_{m}\qpm_{n})=\frac{1}{d}
\ . \label{dmn1}
\end{align}
The following fact follows from the assumptions. The
Hilbert--Schmidt product of two elements from the same POVM can be
described in terms of a single parameter $\varkappa$ \cite{kag14}:
\begin{equation}
\Tr(\ppm_{m}\ppm_{n})=\delta_{mn}{\,}\varkappa
+(1-\delta_{mn}){\>}\frac{1-\varkappa}{d-1}
\ . \label{mjmk}
\end{equation}
General bounds on the parameter $\varkappa$ are written as
$1/d\leq\varkappa\leq1$ \cite{kag14}. The set
$\mpb=\{\ppc^{(1)},\ldots,\ppc^{(M)}\}$ is a set of MUMs of the
efficiency $\varkappa$, when each two measurements obey the above
properties. It turns out that we can reach the aim to build a
complete set of $d+1$ mutually unbiased measurements in $d$
dimensions \cite{kag14}.

Let us consider $d^{2}-1$ operators that form an orthogonal basis
in the space of traceless Hermitian operators. For instance, we
can start with the generators of ${\textup{SU}}(d)$ \cite{kag14}.
Using such operators, one can built a family of traceless
Hermitian operators $\fm_{n}^{(b)}$ with labels $b=1,\ldots,d+1$
and $n=1,\ldots,d$. The constructed operators obey the condition
\begin{equation}
\Tr{\bigl(\fm_{m}^{(a)}\fm_{n}^{(b)}\bigr)}=0
\qquad (a\neq{b})
\ . \label{anqb}
\end{equation}
For one and the same label $b$, they also satisfy \cite{kag14}
\begin{equation}
\Tr{\bigl(\fm_{m}^{(b)}\fm_{n}^{(b)}\bigr)}=\bigl(1+\sqrt{d}\bigr)^{2}
\bigl[\delta_{mn}(d-1)-(1-\delta_{mn})\bigr]
\ , \label{bnqb}
\end{equation}
where $\delta_{mn}$ is the Kronecker symbol.

An explicit construction of MUMs is written as follows
\cite{kag14}. For $b=1,\ldots,d+1$ and $n=1,\ldots,d$, we
introduce operators
\begin{equation}
\ppm_{n}^{(b)}=\frac{\pen}{d}+t{\,}\fm_{n}^{(b)}
\ , \label{pnvfn}
\end{equation}
where $t$ is some parameter that should be chosen. The least
quantities among eigenvalues of the operators $\fm_{n}^{(b)}$
determine an interval, in which $t$ can be varied \cite{kag14}.
This interval should be such that $\ppm_{n}^{(b)}\in\lsp(\hh)$ for
all values of the labels. With the given $t$, the efficiency
parameter is calculated as
\begin{equation}
\varkappa=\frac{1}{d}+t^{2}\bigl(1+\sqrt{d}\bigr)^{2}(d-1)
\ . \label{dvit}
\end{equation}
The range of $t$ leads to the corresponding range of $\varkappa$.
The measurements
$\ppc^{(b)}=\bigl\{\ppm_{n}^{(b)}\bigr\}_{n=1}^{d}$ then form a
complete set of MUMs of the efficiency (\ref{dvit}).

As measures of an uncertainty in quantum measurements, we will use
the R\'{e}nyi and Tsallis entropies. The concept of entropy is of
great importance in both information theory and statistical
physics. In addition to the Shannon entropy, other entropic
measures were found to be useful. The R\'{e}nyi and Tsallis
entropies are both very important \cite{bengtsson}. For the given
probability distribution, its R\'{e}nyi $\alpha$-entropy is
defined as \cite{renyi61}
\begin{equation}
R_{\alpha}(p):=\frac{1}{1-\alpha}{\ }\ln\left(\sum\nolimits_{n} p_{n}^{\alpha}\right)
\, , \label{rpdf}
\end{equation}
where $\alpha>0$ and $\alpha\neq1$. This quantity is a
non-increasing function of $\alpha$ \cite{renyi61}. Other
properties of parametric dependence of (\ref{rpdf}) are discussed
in \cite{zycz}. The Renyi entropy of order $\alpha=2$ is also
known as the collision entropy \cite{ww10,bengtsson}. It is
written as
\begin{equation}
R_{2}(p)={-\ln}{\left(\sum\nolimits_{n}p_{n}^{2}\right)}
\> . \label{clen}
\end{equation}
In the limit $\alpha\to\infty$, we have the so-called min-entropy
\begin{equation}
R_{\infty}(p)=-\ln(\max{p}_{n})
\ . \label{mnen}
\end{equation}
The min-entropy is of specific interest in cryptography
\cite{ngbw12}. It is also linked with the extrema of the discrete
Wigner function \cite{MWB10}. Uncertainty bounds on R\'{e}nyi's
entropies are significant in studying the connection between
complementarity and uncertainty principles \cite{bosyk13a}. Using
the R\'{e}nyi entropy, the writers of \cite{rprz12} obtained
trade-off relations for a trace-preserving quantum operation. An
extension of such trade-off relations in terms of the unified
entropies was given in \cite{rast13a}.

The notion of Tsallis entropy is widely used in non-extensive
statistical mechanics \cite{gmt}. The non-extensive entropy of
positive degree $\alpha\neq1$ is defined as \cite{tsallis}
\begin{equation}
H_{\alpha}(p):=\frac{1}{1-\alpha}{\,}
\left(\sum\nolimits_{n} p_{n}^{\alpha}-1\right)
{\,}. \label{tsent}
\end{equation}
With the factor $\left(2^{1-\alpha}-1\right)^{-1}$ instead of
$(1-\alpha)^{-1}$, this function was deduced by Havrda and
Charv\'{a}t \cite{havrda}. The entropy (\ref{tsent}) is concave
for all $\alpha>0$. In more detail, properties of the entropy
(\ref{tsent}) and related functionals are considered in
\cite{sf06,rastkyb}. It is convenient to rewrite (\ref{tsent}) as
\begin{equation}
H_{\alpha}(p)=-\sum\nolimits_{n}p_{n}^{\alpha}{\,}\ln_{\alpha}(p_{n})
=\sum\nolimits_{n}p_{n}{\>\,}{\ln_{\alpha}}{\left(\frac{1}{p_{n}}\right)}
\ . \label{tsaln}
\end{equation}
Here, we used the $\alpha$-logarithm defined for $\alpha>0\not=1$
and $x>0$ as
\begin{equation}
\ln_{\alpha}(x)=\frac{x^{1-\alpha}-1}{1-\alpha}
\ . \label{lnadf}
\end{equation}
In the limit $\alpha\to1$, we obtain $\ln_{\alpha}(x)\to\ln{x}$
and the standard Shannon entropy
\begin{equation}
H_{1}(p)=-\sum\nolimits_{n}p_{n}{\,}\ln{p}_{n}
\ . \label{shaln}
\end{equation}
Of course, the right-hand side of (\ref{rpdf}) also gives
(\ref{shaln}) in this limit. Various applications of the above
entropies and their quantum counterparts are discussed in the book
\cite{bengtsson}.

Analyzing the case of detection inefficiencies, we will use the
method of \cite{rastmub,rastpsic}. To the given value
$\eta\in[0;1]$ and probability distribution $\{p_{n}\}$, one
assigns a ``distorted'' distribution:
\begin{equation}
p_{n}^{(\eta)}=\eta{\,}p_{n}
\ , \qquad
p_{\varnothing}^{(\eta)}=1-\eta
\ . \label{dspd}
\end{equation}
Here, the parameter $\eta\in[0;1]$ describes a detector
efficiency. By $p_{\varnothing}^{(\eta)}$, we denote the
probability of the no-click event. The described model of
distorted probabilities was introduced in studying entropic Bell
inequalities with detector inefficiencies \cite{rchtf12}. Further
development of Bell inequalities with detection inefficiencies was
given in \cite{rastqqt}. As was shown in the paper \cite{rastqqt},
for all $\alpha>0$ we have
\begin{equation}
{H_{\alpha}}{\bigl(p^{(\eta)}\bigr)}=\eta^{\alpha}H_{\alpha}(p)+h_{\alpha}(\eta)
\ ,  \label{qtlm0}
\end{equation}
where ${H_{\alpha}}{\bigl(p^{(\eta)}\bigr)}$ denotes the entropy
of ``distorted'' distribution (\ref{dspd}). As usual, the binary
Tsallis entropy $h_{\alpha}(\eta)$ is written as
\begin{equation}
h_{\alpha}(\eta):={}-\eta^{\alpha}\ln_{\alpha}(\eta)-(1-\eta)^{\alpha}\ln_{\alpha}(1-\eta)
\ . \label{bnta}
\end{equation}
Entropic uncertainty relations with detection inefficiencies were
derived for mutually unbiased bases in \cite{rastmub} and for a
general SIC-POVM in \cite{rastpsic}. We will apply this method to
mutually unbiased measurements.

Deriving uncertainty relations in terms of the R\'{e}nyi
entropies, we will deal with the following situation. Both the
collision entropy and min-entropy can be calculated or estimated
from below. We wish to obtain a lower bound on the R\'{e}nyi
entropy of order $\alpha\geq2$. An answer to the question is
written as follows.

\newtheorem{prp21}{Proposition}
\begin{prp21}\label{pon21}
For $\alpha\in[2;\infty]$, the R\'{e}nyi $\alpha$-entropy is
bounded from below as
\begin{equation}
R_{\alpha}(p)\geq\frac{1}{\alpha-1}{\>}R_{2}(p)+\frac{\alpha-2}{\alpha-1}{\>}R_{\infty}(p)
\ . \label{al2inf}
\end{equation}
\end{prp21}

{\bf Proof.} We will deal with finite $\alpha\geq2$. We first
write the inequality
\begin{equation}
\sum\nolimits_{n} p_{n}^{\alpha}\leq
(\max{p}_{n})^{\alpha-2}{\,}\sum\nolimits_{n} p_{n}^{2}
\ . \label{ainf1}
\end{equation}
The function $x\mapsto(1-\alpha)^{-1}\ln{x}$ decreases for
$\alpha\geq2$. Combining this fact with the formulas (\ref{rpdf}),
(\ref{clen}), and (\ref{mnen}) completes the proof.
$\blacksquare$

It should be pointed out that, for $\alpha\in[2;\infty]$, the
R\'{e}nyi $\alpha$-entropy can be estimated in terms of only the
collision entropy. As was mentioned in \cite{rastmub}, for
$\alpha\geq2$ we have
\begin{equation}
R_{\alpha}(p)\geq\frac{\alpha}{2(\alpha-1)}{\>}R_{2}(p)
\ . \label{ainf2}
\end{equation}
This fact directly follows from theorem 19 of the book
\cite{hardy}. It is easy to check that the second bound
(\ref{ainf2}) cannot be stronger than (\ref{al2inf}). Subtracting
the right-hand side of (\ref{ainf2}) from the right-hand side of
(\ref{al2inf}), we obtain
\begin{equation}
\frac{\alpha-2}{2(\alpha-1)}
{\,}\Bigl(
2R_{\infty}(p)-R_{2}(p)
\Bigr)\geq0
\ . \label{2rrin}
\end{equation}
The latter is equivalent to
$(\max{p}_{n})^{2}\leq\sum_{n}p_{n}^{2}$. Taking (\ref{al2inf}), we
may generally obtain better bounds. In the following, the result
(\ref{al2inf}) will be used in deriving uncertainty bounds for MUMs
in terms of the R\'{e}nyi entropies.

\section{Indices of coincidence for MUMs}\label{sec3}

In this section, we will study indices of coincidence for
measurements considered. Let $\ppc=\{\ppm_{n}\}_{n=1}^{d}$ be a
mutually unbiased measurement in $d$-dimensional Hilbert space. If
the pre-measurement state is described by density matrix $\bro$,
then the probability on $n$-th outcome is written as
\begin{equation}
p_{n}(\ppc|\bro)=\Tr(\ppm_{n}\bro)
\ . \label{prbn}
\end{equation}
The index of coincidence is then defined as the sum of squared
probabilities, namely
\begin{equation}
C(\ppc|\bro):=\sum_{n=1}^{d} p_{n}(\ppc|\bro)^{2}
\ . \label{icdf}
\end{equation}
It seems to be natural that the sum (\ref{icdf}) can be linked to
the quantity $\Tr(\bro^{2})$. This quantity, called the purity of
$\bro$, is frequently used since it is easy to compute
\cite{bengtsson}. The purity of a quantum state gives a good
characterization of the degree of information about its
preparation \cite{adesso05}. We have the following general bounds
on (\ref{icdf}):
\begin{equation}
\frac{1}{d}\leq{C}(\ppc^{(b)}|\bro)\leq1
\ . \label{1dc1}
\end{equation}
Here, the lower bound follows from the convexity of the function
$x\mapsto{x}^{2}$. The index of coincidence has been exactly
calculated for a single SIC-POVM \cite{rastmub} and for a general
SIC-POVM  \cite{rastpsic}. In both cases, resulting expression
involves purity of the measured state. To derive entropic
uncertainty relations, we wish to estimate from above the sum of
the indices of coincidence for several mutually unbiased
measurements. It is natural that our bound is formulated in terms
of the purity.

\newtheorem{prp31}[prp21]{Proposition}
\begin{prp31}\label{pon31}
Let $\mpb=\{\ppc^{(1)},\ldots,\ppc^{(M)}\}$ be a set of $M$
mutually unbiased measurements of the efficiency $\varkappa$ in
$d$ dimensions. For arbitrary $\bro$, the sum of the corresponding
indices of coincidence obey
\begin{equation}
\sum_{\ppc\in\mpb}C(\ppc|\bro)\leq\frac{M-1}{d}+\frac{1-\varkappa+(\varkappa{d}-1){\,}\Tr(\bro^{2})}{d-1}
\ . \label{ubp1}
\end{equation}
\end{prp31}

{\bf Proof.} It follows from the construction of MUMs that any
density matrix can be represented as \cite{kag14}
\begin{equation}
\bro=\frac{\pen}{d}+\sum_{b=1}^{d+1}\sum_{n=1}^{d} r_{n}^{(b)}{\,}\fm_{n}^{(b)}
\ . \label{bsrp}
\end{equation}
Since the operators $\fm_{n}^{(b)}$ are all traceless and obey
(\ref{anqb}), we have
\begin{equation}
\Tr(\bro^{2})=\frac{1}{d}+\sum_{b=1}^{d+1}\sum_{m,n=1}^{d}
r_{m}^{(b)}r_{n}^{(b)}{\,}\Tr{\bigl(\fm_{m}^{(b)}\fm_{n}^{(b)}\bigr)}
\ . \label{tbr2}
\end{equation}
By (\ref{bnqb}), we have
$\Tr{\bigl(\fm_{m}^{(b)}\fm_{n}^{(b)}\bigr)}=-\bigl(1+\sqrt{d}\bigr)^{2}$
for $m\neq{n}$ and
\begin{equation}
\Tr{\bigl(\fm_{n}^{(b)}\fm_{n}^{(b)}\bigr)}=\bigl(1+\sqrt{d}\bigr)^{2}(d-1)
\ . \label{fnfn}
\end{equation}
Using these formulas, we obtain
\begin{align}
&\sum_{m,n=1}^{d}
r_{m}^{(b)}r_{n}^{(b)}{\,}\Tr{\bigl(\fm_{m}^{(b)}\fm_{n}^{(b)}\bigr)}
\nonumber\\
&=\bigl(1+\sqrt{d}\bigr)^{2}(d-1)\sum_{n=1}^{d}r_{n}^{(b)}r_{n}^{(b)}
-\bigl(1+\sqrt{d}\bigr)^{2}\sum_{\substack{m,n=1 \\ m\neq{n}}}^{d}r_{m}^{(b)}r_{n}^{(b)}
\nonumber\\
&=\bigl(1+\sqrt{d}\bigr)^{2}
\left(d\sum_{n=1}^{d}{r_{n}^{(b)}r_{n}^{(b)}}
-R^{(b)}R^{(b)}
\right)
{\,}, \label{srmrn}
\end{align}
where $R^{(b)}:=\sum_{n=1}^{d}r_{n}^{(b)}$. Substituting
(\ref{srmrn}) into (\ref{tbr2}), we obtain an expression for 
$\Tr(\bro^{2})$.

Combining (\ref{pnvfn}) with (\ref{bsrp}), the probability of
$n$-th outcome in $b$-th measurement reads
\begin{align}
\frac{1}{d}+t{\,}\Tr{\bigl(\bro{\,}\fm_{n}^{(b)}\bigr)}
&=\frac{1}{d}+t\sum_{m=1}^{d}
r_{m}^{(b)}{\,}\Tr{\bigl(\fm_{m}^{(b)}\fm_{n}^{(b)}\bigr)}
\nonumber\\
&=\frac{1}{d}+t\bigl(1+\sqrt{d}\bigr)^{2}
\bigl(d{\,}r_{n}^{(b)}-R^{(b)}
\bigr)
{\,}. \label{pnbp}
\end{align}
The last expression is obtained similarly to (\ref{srmrn}).
Squaring this probability and further summing with respect to
$n=1,\ldots,d$, one gets
\begin{equation}
C(\ppc^{(b)}|\bro)=\frac{1}{d}+t^{2}\bigl(1+\sqrt{d}\bigr)^{4}
\left(d^{2}\sum_{n=1}^{d}{r_{n}^{(b)}r_{n}^{(b)}}
-d{\,}R^{(b)}R^{(b)}
\right)
{\,}. \label{cbbp}
\end{equation}
Here, we used
$\sum_{n=1}^{d}{\bigl(d{\,}r_{n}^{(b)}-R^{(b)}\bigr)}=0$. By
(\ref{1dc1}), we see that, for all $b$, the second term in the
right-hand side of (\ref{cbbp}) is non-negative.

Combining (\ref{tbr2}), (\ref{srmrn}), and (\ref{cbbp}), we
further write
\begin{align}
\sum_{b=1}^{M}C(\ppc^{(b)}|\bro)
&=\frac{M}{d}+t^{2}d\bigl(1+\sqrt{d}\bigr)^{4}\sum_{b=1}^{M}\left(d\sum_{n=1}^{d}{r_{n}^{(b)}r_{n}^{(b)}}
-R^{(b)}R^{(b)}
\right)
\nonumber\\
&\leq\frac{M}{d}+t^{2}d\bigl(1+\sqrt{d}\bigr)^{2}
\left(\Tr(\bro^{2})-\frac{1}{d}\right)
\nonumber\\
&=\frac{M-1}{d}+\varkappa+\frac{\varkappa{d}-1}{d-1}
\left(\Tr(\bro^{2})-1\right)
\, . \label{csmp}
\end{align}
At the last step, we used (\ref{dvit}). The quantity (\ref{csmp})
is easily reduced to the right-hand side of (\ref{ubp1}).
$\blacksquare$

The statement of Proposition \ref{pon31} provides an upper bound
on the sum of the indices of coincidence for a set of MUMs. For
the complete set of $d+1$ MUMs, we actually have an exact result
instead of inequality:
\begin{equation}
\sum_{b=1}^{d+1}C(\ppc^{(b)}|\bro)
=1+\frac{1-\varkappa+(\varkappa{d}-1){\,}\Tr(\bro^{2})}{d-1}
\ . \label{ubpd1}
\end{equation}
Indeed, the inequality (\ref{csmp}) is saturated with $M=d+1$. For
pure states, the right-hand side of (\ref{ubpd1}) becomes
$1+\varkappa$. This result was presented in \cite{kag14} and then
used in the context of entanglement detection in \cite{fei14}. The
inequality (\ref{ubp1}) is tight in the sense that it is always
saturated with the completely mixed state $\bro_{*}=\pen/d$. Since
operators $\fm_{n}^{(b)}$ are all traceless, we see from
(\ref{pnbp}) that
\begin{equation}
p_{n}(\ppc^{(b)}|\bro_{*})=\frac{1}{d}
\ , \label{pnras0}
\end{equation}
irrespectively to $n$ and $b$. For each $b=1,\ldots,d+1$,
therefore, the index of coincidence reads
\begin{equation}
C(\ppc^{(b)}|\bro_{*})=\sum_{n=1}^{d}\frac{1}{d^{2}}=\frac{1}{d}
\ . \label{pnras1}
\end{equation}
Hence, the left-hand side of (\ref{ubp1}) is equal to $M/d$ for
the completely mixed state $\bro_{*}$. By substitution
$\Tr(\bro_{*}^{2})=1/d$, the right-hand side of (\ref{ubp1}) gives
$M/d$ as well. Thus, our result is almost precise for impure
states with the purity close to $1/d$. The bound (\ref{ubp1}) may
also be saturated with pure states. Below, we will shortly mention
an example for MUBs. Note that the purity $\Tr(\bro^{2})$ can be
expressed in terms of the Bloch vector of $\bro$ \cite{rastpsic}.
Hence, the formulas (\ref{ubp1}) and (\ref{ubpd1}) can be
rewritten via the Bloch vector as well. We refrain from presenting
the details here.

It should be noticed that the results (\ref{ubp1}) and
(\ref{ubpd1}) are calculated for the aforementioned construction
of mutually unbiased measurements. Setting $\varkappa=1$,
nevertheless, the inequality (\ref{ubp1}) leads to the correct
result for mutually unbiased bases. As was shown in \cite{molm09},
for a set $\cmb=\{\clb^{(1)},\ldots,\clb^{(M)}\}$ of $M$ mutually
unbiased bases we have
\begin{equation}
\sum_{\clb\in\cmb} C(\clb|\bro)\leq\frac{M-1}{d}+\Tr(\bro^{2})
\ . \label{bap1}
\end{equation}
We consider $\bro=|\psi\rangle\langle\psi|$ with
$|\psi\rangle$ taken from one of the bases
$\clb^{(1)},\ldots,\clb^{(M)}$. In this case, the inequality
(\ref{bap1}) is actually saturated. Indeed, one the indices
$C(\clb|\bro)$ is then equal to $1$ and other are all $1/d$. The
sum of these indices is equal to the right-hand side of
(\ref{bap1}) for $\Tr(\bro^{2})=1$. Using (\ref{bap1}),
uncertainty relations in terms of the Shannon entropies have been
obtained \cite{molm09}. Extensions with the use of generalized
entropies were derived in \cite{rastmub}. The authors of
\cite{shbah12} considered applications of the bound (\ref{bap1})
in the context of entanglement detection. In the following, we
will use (\ref{ubp1}) for obtaining entropic bounds for an
arbitrary set of MUMs. We will also revisit applications of MUMs
in entanglement detection.

\section{Entropic uncertainty relations for MUMs}\label{sec4}

In this section, we present uncertainty relations for an arbitrary
set of mutually unbiased measurements. First, we obtain
uncertainty relations in terms of the R\'{e}nyi entropies. Second,
we give formulations in terms of the Tsallis $\alpha$-entropies of
order $\alpha\in(0;2]$. Both the state-dependent and
state-independent formulations are considered. In the Tsallis
case, we also address uncertainty relations with detection
inefficiencies. Our first result is posed as follows.

\newtheorem{prp41}[prp21]{Proposition}
\begin{prp41}\label{pon41}
Let $\mpb=\{\ppc^{(1)},\ldots,\ppc^{(M)}\}$ be a set of $M$
mutually unbiased measurements of the efficiency $\varkappa$ in
$d$ dimensions. For $\alpha\in[2;\infty]$ and arbitrary density
matrix $\bro$ on $\hh$, the averaged sum of R\'{e}nyi's entropies satisfies
the state-dependent bound
\begin{align}
\frac{1}{M}{\,}\sum_{\ppc\in\mpb} R_{\alpha}(\ppc|\bro)
&\geq\frac{1}{\alpha-1}{\>}
{\ln}{\left(\frac{Md(d-1)}{M(d-1)+(\varkappa{d}-1)\bigl(\Tr(\bro^{2})d-1\bigr)}\right)}
\label{rum01}\\
&+\frac{\alpha-2}{\alpha-1}
{\,}\biggl\{
\ln{d}-{\ln}{\left(1+M^{-1/2}\sqrt{\varkappa{d}-1}{\,}\sqrt{\Tr(\bro^{2})d-1}\right)}
\biggr\}
{\>}. \nonumber
\end{align}
\end{prp41}

{\bf Proof.} First, we will obtain a lower bound on the averaged sum of
collision entropies. Since the function $x\mapsto-\ln{x}$ is
convex, we write
\begin{equation}
\frac{1}{M}{\,}\sum_{\ppc\in\mpb} R_{2}(\ppc|\bro)=
\sum_{\ppc\in\mpb}\frac{1}{M}{\,}\Bigl(-\ln{C}(\ppc|\bro)\Bigr)
\geq{-\ln}{\left(\frac{1}{M}{\,}\sum_{\ppc\in\mpb}C(\ppc|\bro)\right)}
{\>}. \label{cov2}
\end{equation}
It follows from (\ref{ubp1}) that
\begin{equation}
\frac{1}{M}{\,}\sum_{\ppc\in\mpb}C(\ppc|\bro)\leq
\frac{M(d-1)+(\varkappa{d}-1)\bigl(\Tr(\bro^{2})d-1\bigr)}{Md(d-1)}
\ . \label{ubpm1}
\end{equation}
As the function $x\mapsto-\ln{x}$ decreases, combining
(\ref{cov2}) with (\ref{ubpm1}) leads to the result
\begin{equation}
\frac{1}{M}{\,}\sum_{\ppc\in\mpb} R_{2}(\ppc|\bro)\geq
{\ln}{\left(\frac{Md(d-1)}{M(d-1)+(\varkappa{d}-1)\bigl(\Tr(\bro^{2})d-1\bigr)}\right)}
{\>}. \label{cov22}
\end{equation}
The second step is to get a lower bound on the averaged sum of
min-entropies. It follows from lemma 3 of \cite{rastmub} that
\begin{equation}
\max\bigl\{p_{n}:{\,}1\leq{n}\leq{d}\bigr\}
\leq\frac{1}{d}\left(1+\sqrt{d-1}{\,}\sqrt{C(p)d-1}\right)
{\,}, \label{aust}
\end{equation}
where $C(p)$ is the index of coincidence. For clarity, we
introduce the function
\begin{equation}
g_{d}(x):=\frac{1}{d}{\>}\Bigl(1+\sqrt{d-1}{\,}\sqrt{xd-1}\Bigr)
{\>}. \label{gdxf}
\end{equation}
This function is concave and increasing. Combining these facts
with (\ref{aust}) and (\ref{ubpm1}), we obtain
\begin{align}
&\frac{1}{M}{\,}\sum_{\ppc\in\mpb}
\underset{n}{\max}{\,}p_{n}(\ppc|\bro)
\leq
\sum_{\ppc\in\mpb}\frac{1}{M}{\>\,}{g_{d}}{\bigl(C(\ppc|\bro)\bigr)}
\leq
g_{d}{\left(\frac{1}{M}{\,}\sum_{\ppc\in\mpb} C(\ppc|\bro)\right)}
\nonumber\\
&\leq{g}_{d}{\left(
\frac{M(d-1)+(\varkappa{d}-1)\bigl(\Tr(\bro^{2})d-1\bigr)}{Md(d-1)}
\right)}
{\>}. \label{grp2}
\end{align}
Calculating the term (\ref{grp2}) in line with the definition
(\ref{gdxf}) leads to
\begin{equation}
\frac{1}{M}{\,}\sum_{\ppc\in\mpb}
\underset{n}{\max}{\,}p_{n}(\ppc|\bro)
\leq\frac{1}{d}{\,}
\left(1+M^{-1/2}\sqrt{\varkappa{d}-1}{\,}\sqrt{\Tr(\bro^{2})d-1}\right)
\, . \label{res1}
\end{equation}
Due to convexity of the function $x\mapsto-\ln{x}$, we further
write
\begin{equation}
\frac{1}{M}{\,}\sum_{\ppc\in\mpb} R_{\infty}(\ppc|\bro)\geq
{-\ln}{\left(
\frac{1}{M}{\,}\sum_{\ppc\in\mpb}
\underset{n}{\max}{\,}p_{n}(\ppc|\bro)
\right)}
\> . \label{res2}
\end{equation}
Since this function decreases, combining (\ref{res1}) with
(\ref{res2}) finally gives
\begin{equation}
\frac{1}{M}{\,}\sum_{\ppc\in\mpb} R_{\infty}(\ppc|\bro)\geq
\ln{d}-{\ln}{\left(1+M^{-1/2}\sqrt{\varkappa{d}-1}{\,}\sqrt{\Tr(\bro^{2})d-1}\right)}
\> . \label{res3}
\end{equation}
Using both the bounds (\ref{cov22}) and (\ref{res3}), we complete
the proof of (\ref{rum01}) due to (\ref{al2inf}).
$\blacksquare$

The statement of Proposition \ref{pon41} gives a state-dependent
lower bound on the sum of corresponding R\'{e}nyi's entropies. In
the case of MUBs, we have $\varkappa=1$. Let
$\cmb=\{\clb^{(1)},\ldots,\clb^{(M)}\}$ be a set of $M$ mutually
unbiased bases. For $\alpha\in[2;\infty]$, there holds
\begin{align}
\frac{1}{M}{\,}\sum_{\clb\in\cmb} R_{\alpha}(\clb|\bro)
&\geq\frac{1}{\alpha-1}{\>}
{\ln}{\left(\frac{Md}{M+\Tr(\bro^{2})d-1}\right)}
\label{rum01b}\\
&+\frac{\alpha-2}{\alpha-1}
{\,}\biggl\{
\ln{d}-{\ln}{\left(1+M^{-1/2}\sqrt{d-1}{\,}\sqrt{\Tr(\bro^{2})d-1}\right)}
\biggr\}
{\>}. \nonumber
\end{align}
For internal points of the interval $\alpha\in[2;\infty]$, this
result gives an improvement of the uncertainty relations of
\cite{rastmub}.

The bound (\ref{rum01}) is tight in the following sense. For
arbitrary $\alpha\in[2;\infty]$, this inequality is certainly
saturated with the completely mixed state. Substituting
$\Tr(\bro_{*}^{2})=1/d$, for $\alpha\geq2$ the relation
(\ref{rum01}) finally gives
\begin{equation}
\frac{1}{M}{\,}\sum_{\ppc\in\mpb} R_{\alpha}(\ppc|\bro_{*})\geq
\frac{1}{\alpha-1}{\>}\ln{d}+\frac{\alpha-2}{\alpha-1}{\>}\ln{d}=\ln{d}
\ . \label{exrp}
\end{equation}
Since the distribution (\ref{pnras0}) is uniform, we have
$R_{\alpha}(\ppc|\bro_{*})=\ln{d}$ for all $\alpha>0$. Thus, the
lower bound is actually saturated.

In the case of pure states, we obtain a state-independent lower
bound. If $\bro=|\psi\rangle\langle\psi|$, then we have
$\Tr(\bro^{2})=1$. Substituting this into the right-hand side of
(\ref{rum01}), one gets
\begin{align}
\frac{1}{M}{\,}\sum_{\ppc\in\mpb} R_{\alpha}(\ppc|\psi)
&\geq\frac{1}{\alpha-1}{\>}
{\ln}{\left(\frac{Md}{M+\varkappa{d}-1}\right)}
\label{rum01p}\\
&+\frac{\alpha-2}{\alpha-1}
{\,}\biggl\{
\ln{d}-{\ln}{\left(1+M^{-1/2}\sqrt{\varkappa{d}-1}\sqrt{d-1}\right)}
\biggr\}
{\>}. \nonumber
\end{align}
Of course, this lower bound is also valid for all mixed states.
For impure $\bro$, the lower bound (\ref{rum01}) is stronger than
(\ref{rum01p}) due to $\Tr(\bro^{2})<1$. That is, the right-hand
side of (\ref{rum01}) increases with a deviation of the purity
from $1$. Dependence of such a kind seems to be natural.

The bound (\ref{rum01}) covers the interval $\alpha\in[2;\infty]$.
For the interval $\alpha\in(0;2]$, we have a lower bound
independent of $\alpha$. Recall that the R\'{e}nyi entropy cannot
increase with growth of $\alpha$. When $0<\alpha<2$, the formula
(\ref{cov22}) remains valid after replacing $R_{2}(\ppc|\bro)$
with $R_{\alpha}(\ppc|\bro)$. In particular, we obtain lower
bounds on the sum of Shannon entropies. For a set $\mpb$ of $M$
MUMs of the efficiency $\varkappa$, we have
\begin{align}
\frac{1}{M}{\,}\sum_{\ppc\in\mpb} H_{1}(\ppc|\bro)
&\geq
{\ln}{\left(\frac{Md(d-1)}{M(d-1)+(\varkappa{d}-1)\bigl(\Tr(\bro^{2})d-1\bigr)}\right)}
\nonumber\\
&\geq{\ln}{\left(\frac{Md}{M+\varkappa{d}-1}\right)}
{\>}. \label{smum02}
\end{align}
Substituting $M=d+1$, the latter formula gives the
state-independent relation derived in \cite{kag14}. Thus, we have
extended this result in the following three directions: (i) our
bounds hold for any set of MUMs; (ii) they are written in terms of
generalized entropies; (iii) they are state-dependent. Let us
proceed to the Tsallis formulation.

\newtheorem{prp42}[prp21]{Proposition}
\begin{prp42}\label{pon42}
Let $\mpb=\{\ppc^{(1)},\ldots,\ppc^{(M)}\}$ be a set of $M$
mutually unbiased measurements of the efficiency $\varkappa$ in
$d$ dimensions. For $\alpha\in(0;2]$ and arbitrary density matrix
$\bro$ on $\hh$, the averaged sum of Tsallis' entropies satisfies the
state-dependent bound
\begin{equation}
\frac{1}{M}{\,}\sum_{\ppc\in\mpb} H_{\alpha}(\ppc|\bro)\geq
{\ln_{\alpha}}{\left(\frac{Md(d-1)}{M(d-1)+(\varkappa{d}-1)\bigl(\Tr(\bro^{2})d-1\bigr)}\right)}
{\>}. \label{mum02}
\end{equation}
\end{prp42}

{\bf Proof.} We will use the following fact. For $\alpha\in(0;2]$
and arbitrary probability distribution, the Tsallis
$\alpha$-entropy obeys \cite{rastmub}
\begin{equation}
H_{\alpha}(p)\geq{\ln_{\alpha}}{\left(\frac{1}{C(p)}\right)}
{\,}. \label{cnv1}
\end{equation}
The inequality (\ref{cnv1}) is based on the fact that the function
$x\mapsto\ln_{\alpha}(1/x)$ is convex for $\alpha\in(0;2]$.
Applying Jensen's inequality to this convex function, one gets
\begin{equation}
\frac{1}{M}{\,}\sum_{\ppc\in\mpb} H_{\alpha}(\ppc|\bro)\geq
\sum_{\ppc\in\mpb}\frac{1}{M}{\>}{\ln_{\alpha}}{\left(\frac{1}{C(\ppc|\bro)}\right)}
\geq{\ln_{\alpha}}{\left\{\biggl(\frac{1}{M}{\,}\sum_{\ppc\in\mpb}C(\ppc|\bro)\biggr)^{{\!}-1}\right\}}
{\>}. \label{cnv2}
\end{equation}
From $x\leq{y}$ we have $\ln_{\alpha}(1/x)\geq\ln_{\alpha}(1/y)$,
since the function $x\mapsto\ln_{\alpha}(1/x)$ is decreasing.
Combining (\ref{cnv2}) with (\ref{ubpm1}) completes the proof.
$\blacksquare$

The statement of Proposition \ref{pon42} gives a state-dependent
lower bound on the average Tsallis entropy for a set of $M$ MUMs.
Setting $\varkappa=1$, the bound (\ref{mum02}) is reduced to the
uncertainty relation derived in \cite{rastmub} for mutually
unbiased bases. For a pure state $\bro=|\psi\rangle\langle\psi|$,
its purity is equal to $1$. Then the bound (\ref{mum02}) reads
\begin{equation}
\frac{1}{M}{\,}\sum_{\ppc\in\mpb} H_{\alpha}(\ppc|\psi)\geq
{\ln_{\alpha}}{\left(\frac{Md}{M+\varkappa{d}-1}\right)}
{\>}. \label{mum02pur}
\end{equation}
It is easy to see that the state-independent lower bound
(\ref{mum02pur}) remain valid for all states. Similarly to
(\ref{rum01}), the right-hand side of (\ref{mum02}) increases as
the purity decreases. When $\alpha=1$, the formulas (\ref{mum02})
and (\ref{mum02pur}) also lead to (\ref{smum02}).

Let us consider uncertainty relations with detection
inefficiencies. By the parameter $\eta\in[0;1]$, we characterize
an efficiency of used detectors. The maximum $\eta=1$ corresponds
to the inefficiency-free case. We will assume that, for any MUM,
the inefficiency-free distribution is distorted according to
(\ref{dspd}). In other words, for all $\ppc\in\mpb$ we write
\begin{equation}
p_{n}^{(\eta)}(\ppc|\bro)=\eta{\,}p_{n}(\ppc|\bro)
\ , \qquad
p_{\varnothing}^{(\eta)}(\ppc|\bro)=1-\eta
\ . \label{dsmd1}
\end{equation}
By $H_{\alpha}^{(\eta)}(\ppc|\bro)$, we mean the $\alpha$-entropy
calculated for (\ref{dsmd1}). The theoretical value
$H_{\alpha}(\ppc|\bro)$ is related to the case of
inefficiency-free implementation of measurements. Using
(\ref{qtlm0}) and (\ref{mum02}), we obtain
\begin{equation}
\frac{1}{M}{\,}\sum_{\ppc\in\mpb} H_{\alpha}^{(\eta)}(\ppc|\bro)\geq
\eta^{\alpha}{\>}{\ln_{\alpha}}{\left(\frac{Md(d-1)}{M(d-1)+(\varkappa{d}-1)\bigl(\Tr(\bro^{2})d-1\bigr)}\right)}
+h_{\alpha}(\eta)
\ , \label{mum2et}
\end{equation}
where $\alpha\in(0;2]$ and $\mpb$ is a set of $M$ MUMs of the
efficiency $\varkappa$. The result (\ref{mum2et}) is an entropic
uncertainty relation in the model of detection inefficiencies. We
see that the inefficiency-free bound (\ref{mum02}) is multiplied
by the factor $\eta^{\alpha}$ and also added by the binary entropy
$h_{\alpha}(\eta)$. Thus, additional uncertainties are induced by
non-ideal detectors \cite{rastmub,rastpsic}. By setting
$\alpha=1$, the relation (\ref{mum2et}) gives a lower bound on the
sum of the Shannon entropies.

In their usual form, uncertainty relations are not directly
applicable to study cryptographic security. To fill such a gap,
entropic uncertainty relations in the presence of quantum memory
should be considered \cite{renes10,coles12,piani14}. In the case
of two measurements, entropic uncertainty relations can be based
on majorization techniques. This important approach has been
studied in recent works \cite{prz13,fgg13,rpz14}. In certain
cases, the majorization approach has allowed to improve previous
bounds. It would be interesting to apply majorization techniques
to mutually unbiased measurement and compare resulting bounds with
the above one. We hope to address this question in future
investigations.

\section{Entanglement detection with arbitrary set of MUMs}\label{sec5}

In this section, we address a problem of entanglement detection.
Applications of mutually unbiased bases in this question were
analyzed in \cite{shbah12}. The entanglement detection via
SIC-POVMs has shortly been discussed in \cite{rastmub}. The use
of a general SIC-POVM for such purposes was discussed in
\cite{lifei14}. The authors of \cite{fei14} extended some results
of \cite{shbah12} with the use of mutually unbiased measurements.
More separability criteria are discussed in \cite{zhsl98,zhao14}.
Note that the scheme of \cite{fei14} is very particular in the
sense that it is based on using only complete sets of MUMs. Hence,
this scheme needs $d(d+1)$ local POVM elements. At the same time,
the method based on a SIC-POVM uses $d^{2}$ POVM elements
\cite{lifei14}. Here, we deal with a number of measurement
operators to be performed. From the viewpoint of implementation,
this number may be treated as an experimental complexity of the
given scheme of entanglement detection. We will show that
implementation costs of entanglement detection with MUMs can be
reduced essentially.

We now consider a bipartite system of two $d$-dimensional
subsystems. Its Hilbert space is the product
$\hh_{AB}=\hh_{A}\otimes\hh_{B}$ of two isomorphic spaces
$\hh_{A}$ and $\hh_{B}$. Let us choose the orthonormal basis
$\bigl\{|i_{S}\rangle\bigr\}$, where $S=A,B$, for each of the two
spaces $\hh_{A}$ and $\hh_{B}$. A maximally entangled pure state
is then expressed as
\begin{equation}
|\Phi_{AB}^{+}\rangle=\frac{1}{\sqrt{d}}{\,}\sum_{i=1}^{d} |i_{A}\rangle\otimes|i_{B}\rangle
\ . \label{mest}
\end{equation}
Entangled states are a basic resource in quantum information
science. Hence, the problem of efficient detection of entanglement
is of great importance \cite{hhhh09,mhph1999}.

Let us recall shortly basic notions. A product state is any state
of the form $\bro_{A}\otimes\bro_{B}$ \cite{bengtsson}. When both
the matrices $\bro_{A}$ and $\bro_{B}$ are rank-one, we have a
pure product state. A bipartite mixed state is called separable,
when its density matrix $\wbro_{AB}$ can be represented as a
convex combination of product states \cite{zhsl98}. That is, there
exist a probability distribution $\{q_{k}\}$ and two sets
$\{\bro_{A}^{(k)}\}$ and $\{\bro_{B}^{(k)}\}$ such that
\begin{equation}
\wbro_{AB}=\sum\nolimits_{k} q_{k}{\>}\bro_{A}^{(k)}\otimes\bro_{B}^{(k)}
\ . \label{sepsdf}
\end{equation}
Note that each separable state can also be expressed as a convex
combination of only pure product states. This fact easily follows
from (\ref{sepsdf}) by substitution of the corresponding spectral
decompositions. When representations of the form (\ref{sepsdf}) is
not possible, the state is called entangled \cite{zhsl98}.

To detect entanglement, we aim to use a collection of local
measurements. Let
$\mpb_{A}=\{\ppc_{A}^{(1)},\ldots,\ppc_{A}^{(M)}\}$ and
$\mqb_{B}=\{\qpc_{B}^{(1)},\ldots,\qpc_{B}^{(M)}\}$ be two sets of
$M$ MUMs. In $b$-th joint measurement, the pair $(m,n)$ of local
outcomes occurs with the probability
\begin{equation}
P^{(b)}(m,n)=
{\Tr}{\left(\bigl(\ppm_{m}^{(b)}\otimes\qpm_{n}^{(b)}\bigr)\wbro_{AB}\right)}
\, . \label{plmn}
\end{equation}
Following the idea of \cite{shbah12}, we introduce a quantity
\begin{equation}
J_{M}(\wbro_{AB})=\sum_{b=1}^{M}\sum_{n=1}^{d}
P^{(b)}(n,n)
\ . \label{jmnbn}
\end{equation}
In the case of mutually unbiased bases, this correlation measure
was proposed in \cite{shbah12}. The definition (\ref{jmnbn}) is an
immediate extension to MUMs.

\newtheorem{prp51}[prp21]{Proposition}
\begin{prp51}\label{pon51}
Let $\mpb_{A}$ be a set of $M$ mutually unbiased measurements of
the efficiency $\varkappa_{A}$ in the $d$-dimensional space
$\hh_{A}$. Let $\mqb_{B}$ be a set of $M$ mutually unbiased
measurements of the efficiency $\varkappa_{B}$ in the
$d$-dimensional space $\hh_{B}$. For all density matrices
$\bro_{A}\in\lsp(\hh_{A})$ and $\bro_{B}\in\lsp(\hh_{B})$, the
correlation measure satisfies
\begin{equation}
J_{M}(\bro_{A}\otimes\bro_{B})\leq
\prod_{S=A,B}
{\left(\frac{M(d-1)+(\varkappa_{S}d-1)\bigl(\Tr(\bro_{S}^{2})d-1\bigr)}{d(d-1)}\right)}^{1/2}
\, . \label{jrrab}
\end{equation}
\end{prp51}

{\bf Proof.} For a product state
$\wbro_{AB}=\bro_{A}\otimes\bro_{B}$, we clearly have
\begin{equation}
P^{(b)}(n,n)=
{p_{n}}{\bigl(\ppc_{A}^{(b)}|\bro_{A}\bigr)}{\,}
{p_{n}}{\bigl(\qpc_{B}^{(b)}|\bro_{B}\bigr)}
\>. \label{pnnab}
\end{equation}
Using equation (\ref{pnnab}) and the Cauchy--Schwarz inequality,
we then obtain
\begin{align}
J_{M}(\bro_{A}\otimes\bro_{B})&=
\sum_{b=1}^{M}\sum_{n=1}^{d}
{p_{n}}{\bigl(\ppc_{A}^{(b)}|\bro_{A}\bigr)}{\,}
{p_{n}}{\bigl(\qpc_{B}^{(b)}|\bro_{B}\bigr)}
\nonumber\\
&\leq
{\left(
\sum_{b=1}^{M}{C}{\bigl(\ppc_{A}^{(b)}|\bro_{A}\bigr)}
\right)^{1/2}}
{\left(
\sum_{b=1}^{M}{C}{\bigl(\qpc_{B}^{(b)}|\bro_{B}\bigr)}
\right)^{1/2}}
\, . \label{cnab}
\end{align}
Combining the latter with (\ref{ubp1}) completes the proof.
$\blacksquare$

The statement of Proposition \ref{pon51} leads to a necessary
criterion that the given bipartite state is a product state.
Applying this criterion for the given input $\wbro_{AB}$, we
should treat $\bro_{A}$ and $\bro_{B}$ as the reduced density
matrices. As usual, they are obtained by the partial trace
operation:
\begin{equation}
\bro_{A}=\Tr_{B}(\wbro_{AB})
\ , \qquad
\bro_{B}=\Tr_{A}(\wbro_{AB})
\ . \label{rarb}
\end{equation}
Relations between some norms of operators before and after partial
trace with applications to quantum entropies were obtained in
\cite{rastjst12}. To use the formula (\ref{jrrab}), we need only
purities of these density matrices. Substituting the purities in
the right-hand side of (\ref{jrrab}), we should then compare the
result with the actual value $J_{M}(\wbro_{AB})$. This value is
calculated from the measurement statistics. If the condition
(\ref{jrrab}) is violated, then the input $\wbro_{AB}$ is
certainly not a product state.

It is of interest to adopt (\ref{jrrab}) for the case, when
purities of the reduced density matrices are unknown. If we keep
at least purity of the state $\wbro_{AB}$ {\it per se}, then a
state-dependent form of the criterion can still be given. If the
input $\wbro_{AB}$ is a product state then
\begin{equation}
\Tr(\bro_{A}^{2}){\,}\Tr(\bro_{B}^{2})=\Tr(\wbro_{AB}^{2})
\ , \qquad
\Tr(\bro_{A}^{2})+\Tr(\bro_{B}^{2})\leq1+\Tr(\wbro_{AB}^{2})
\ . \label{arrb}
\end{equation}
The second relation is proved as follows. If real numbers
$x,y\in[0;1]$ are connected as $xy=a$, then $x+y\leq1+a$, by
convexity of the function $x\mapsto{x}+a/x$. Combining
(\ref{jrrab}) with (\ref{arrb}) leads to the following statement.
If the given density matrix $\wbro_{AB}$ is a product then
\begin{equation}
J_{M}(\wbro_{AB})\leq
\frac{1}{d(d-1)}{\>}
\sqrt{\Gamma^{2}+(\varkappa{d}-1)\bigl[\Gamma{d}
+(M+\varkappa{d}-1)(d-1)d{\,}\Tr(\wbro_{AB}^{2})\bigr]}
\ , \label{jabr1}
\end{equation}
where $\Gamma=M(d-1)-(\varkappa{d}-1)$. The formula (\ref{jabr1})
gives a necessary criterion, which depends on purity of the tested
bipartite state. As the violation of (\ref{jabr1}) has been
observed, we truly conclude that the tested state is not a product
state. Finally, we present the following state-independent bound.

\newtheorem{prp52}[prp21]{Proposition}
\begin{prp52}\label{pon52}
Let $\mpb_{A}$ be a set of $M$ mutually unbiased measurements of
the efficiency $\varkappa_{A}$ in the $d$-dimensional space
$\hh_{A}$. Let $\mqb_{B}$ be a set of $M$ mutually unbiased
measurements of the efficiency $\varkappa_{B}$ in the
$d$-dimensional space $\hh_{B}$. If the given state
$\wbro_{AB}$ is separable then
\begin{equation}
J_{M}(\wbro_{AB})\leq
\frac{1}{d}{\,}\sqrt{M+\varkappa_{A}d-1}{\,}\sqrt{M+\varkappa_{B}d-1}
\ . \label{jabr52}
\end{equation}
\end{prp52}

{\bf Proof.} We first observe the following. For any product
state, the quantity $J_{M}(\bro_{A}\otimes\bro_{B})$ is bounded
from above by the right-hand side of (\ref{jabr52}). This claim
directly follows from (\ref{jrrab}) and $\Tr(\bro_{S}^{2})\leq1$.
As the function $x\mapsto{x}^{2}$ is convex, we further have
\begin{equation}
J_{M}(\wbro_{AB})\leq\sum\nolimits_{k} q_{k}{\>}{J_{M}}{\bigl(\bro_{A}^{(k)}\otimes\bro_{B}^{(k)}\bigr)}
\> . \label{lkkab}
\end{equation}
Combining (\ref{lkkab}) with the above observation completes the
proof due to the condition $\sum_{k}q_{k}=1$. $\blacksquare$

The statement of Proposition \ref{pon52} gives a necessary
criterion for separability of bipartite states. In the case
$\varkappa_{A}=\varkappa_{B}=\varkappa$, this criterion reads
\begin{equation}
J_{M}(\wbro_{AB})\leq
\frac{M+\varkappa{d}-1}{d}
\ . \label{jabr2}
\end{equation}
With $M=d+1$, we deal with the scheme
using two complete sets of MUMs. Then the formula (\ref{jabr2}) is
reduced to $J_{M}(\wbro_{AB})\leq1+\varkappa$. The
latter necessary criterion for the separability was discussed in \cite{fei14}. We have seen that
entanglement detection may be proceeded with a lesser number of
measurements. This approach could be easy for implementation. On
the other hand, a sensitivity of the scheme will probably
decrease. In the following, we give a reason for trade-off between
a sensitivity of the scheme and its experimental complexity.

In general, the above upper bounds give only a necessary criterion
for the separability of bipartite states. For some classes of
states, however, this criterion may be sufficient as well. We
shall now illustrate this fact with isotropic states. Recall that
isotropic states are states of the form
\begin{equation}
\wbro_{{iso}}
=\gamma{\,}|\Phi_{AB}^{+}\rangle\langle\Phi_{AB}^{+}|+(1-\gamma){\,}\wbro_{AB*}
\ . \label{isost}
\end{equation}
Here, $\gamma\in[0;1]$ and $\wbro_{AB*}$ is the completely mixed
state on $\hh_{A}\otimes\hh_{B}$, namely
\begin{equation}
\wbro_{AB*}=\frac{\pen_{A}\otimes\pen_{B}}{d^{2}}
\ . \label{cmsab}
\end{equation}
To the given MUM $\ppc_{A}^{(b)}=\{\ppm_{n}\}$ of the efficiency
$\varkappa$, we assign a set $\ppc_{B}^{(b)*}=\{\ppm_{n}^{*}\}$ of
operators that are conjugate in the following sense. For all
$i,j=1,\ldots,d$, matrix elements obey
\begin{equation}
\langle{i}_{B}|{\,}\ppm_{n}^{*}|j_{B}\rangle=
\langle{j}_{A}|{\,}\ppm_{n}|i_{A}\rangle
\ . \label{cjpas}
\end{equation}
It is easy to check that the set $\ppc_{B}^{(b)*}$ is also a MUM
of the efficiency $\varkappa$. Using the property (\ref{cjpas}),
we obtain
\begin{equation}
\langle\Phi_{AB}^{+}|{\,}\ppm_{n}\otimes\ppm_{n}^{*}|\Phi_{AB}^{+}\rangle
=\frac{1}{d}{\>}\Tr(\ppm_{n}\ppm_{n})=\frac{\varkappa}{d}
\ . \label{apbp}
\end{equation}
For the isotropic state (\ref{isost}), calculations then lead to
the result
\begin{equation}
J_{M}(\wbro_{{iso}})=M
{\left(
\gamma\varkappa+\frac{1-\gamma}{d}
\right)}
\, . \label{jmiso}
\end{equation}
For the value $M=d+1$, this result was presented in \cite{fei14}.
Using the scheme with $M$ MUMs, we can certainly detect
entanglement for those values $\gamma$ that satisfy
\begin{equation}
\frac{M+\varkappa{d}-1}{Md}<\gamma\varkappa+\frac{1-\gamma}{d}
\ , \label{cogam}
\end{equation}
or merely $\gamma>1/M$. Thus, schemes with two sets of $M$
mutually unbiased measurements allow to detect the entanglement of
all isotropic states with $\gamma\in(1/M;1]$. For schemes with
mutually unbiased bases, this result was discussed in
\cite{shbah12}. In this sense, there is a good reason to realize
entanglement detection with MUMs instead of MUBs. The interval
$\gamma\in(1/M;1]$ widens with growth of $M$. Here, we see
some trade-offs between a sensitivity of the considered scheme and
costs for its implementation. The maximal interval is obtained for
$M=d+1$. It is known that isotropic states are certainly entangled
for $\gamma>1/(d+1)$ \cite{mhph1999}. Hence, using two complete
sets of MUMs allows to detect all the entanglement of isotropic
states \cite{fei14}.

Note that the described scheme of entanglement detection can be
realized for arbitrary $d$. Indeed, the explicit construction for
$d+1$ mutually unbiased measurements has been presented in
\cite{kag14}. However, acceptable values of $\varkappa$ cannot be
chosen {\it a priori}. In this way, therefore, we cannot generally
obtain a set of $d+1$ mutually unbiased bases. The problem of
maximal set of mutually unbiased bases seems to be very hard. In
practical questions, however, we may try to adopt MUMs instead of
MUBs. This possibility was already discussed in
\cite{kag14,fei14}. The above results also support such an
approach to applications of mutual unbiasedness in quantum
information processing.

\section{Conclusions}\label{sec6}

We have studied some  properties of recently introduced mutually
unbiased measurements. For a set of several MUMs, we estimated
from above the sum of corresponding indices of coincidence. The
presented results are essentially based on such calculations. Then
we have obtained entropic uncertainty relations for a set of
several MUMs in terms of the R\'{e}nyi and Tsallis entropies. The
derived relations include both the state-dependent and
state-independent forms. The state-dependent bound on the sum of
corresponding R\'{e}nyi's entropies turns to be tight in some
sense. Namely, for all orders this bound is saturated with the
completely mixed state. We also obtained some improvement of the
R\'{e}nyi entropic bounds for mutually unbiased bases. The Tsallis
formulation allowed to address uncertainty relations with
detection inefficiencies. Applications of mutually unbiased
measurements in entanglement detection were considered in more
details. In particular, we obtained results for an arbitrary
number of MUMs used in entanglement detection. It seems that there
is a certain trade-off between a sensitivity of the scheme and
costs for its experimental implementation. In the literature,
mutually unbiased bases were considered as a suitable tool in
quantum state reconstruction, quantum error correction, and the
mean king's problem. It would be interesting to study possible
applications of MUMs in these questions. In this regard, the
results presented here may also be of significance.

\medskip

{\it Note added.} After this work was completed I learned about very recent results of Chen and Fei \cite{chenfei14}. These authors independently studied uncertainty relations for MUMs in terms of the R\'{e}nyi and Tsallis entropies. The uncertainty relations presented in my work differ in the following two respects. First, an arbitrary number of MUMs was considered, whereas the authors of \cite{chenfei14} deal with $d+1$ MUMs. Second, for $M=d+1$ the lower bound (\ref{rum01}) is stronger than the corresponding bound of \cite{chenfei14}.

\medskip

The author acknowledges fruitful discussions with Zbigniew
Pucha{\l}a and Karol \.{Z}yczkowski in Institute of Physics,
Jagiellonian University, Krak\'{o}w.

\end{document}